\documentclass[11pt,preprint]{aastex}
\shorttitle{Gradients in M32}
\shortauthors{Guy Worthey}

\begin{document}

\title{Balmer and Metal Absorption Feature Gradients in M32}

\author{Guy Worthey}
\affil{Program in Astronomy, Washington State University, Pullman, WA
99164-2814}

\begin{abstract}

New data sources are used to assess Lick/IDS feature strength
gradients inside the half-light radius $R_e$ of the compact Local
Group elliptical galaxy M32. A {\em Hubble Space Telescope} (HST) STIS
spectrum seemed to indicate ionized gas and a very young central
stellar population. In fact, this conclusion is entirely spurious
because of incomplete removal of ion hits. More robust ground-based
spectra taken at the MDM Observatory are, in contrast, the most
accurate measurements of Lick/IDS indices yet obtained for M32.  All
but a few (of 24 measured) indices show a statistically significant
gradient.  The CN indices show a maximum at 4\arcsec\ radius, dropping
off both toward the nucleus and away from it.  At 2\arcsec\ radius
there is a discontinuity in the surface brightness profile, but this
feature is not reflected in any spectral feature.  Comparing with
models, the index gradients indicate a mean age and abundance gradient
in the sense that the nucleus is a factor of 2.5 younger and a factor
of 0.3 dex more metal-rich than at 1 $R_e$. This conclusion is only
weakly dependent on which index combinations are used and is robust to
high accuracy.  Stars near the M32 nucleus have a mean age and heavy
element abundance [M/H] of (4.7 Gyr, +0.02), judging from models by
Worthey with variable abundance ratios. This result has very small
formal random errors, although, of course, there is significant
age-metallicity degeneracy along an (age, abundance) line segment from
(5.0 Gyr, 0.00) to (4.5 Gyr, +0.05). An abundance pattern of
[C/M]$=+0.077$ (carbon abundance affects CN, C$_2$4668, and the bluer
Balmer features), [N/M]$=-0.13$, [Mg/M]$=-0.18$, [Fe/M]$\approx$0.0,
and [Na/M]$=+0.12$ is required to fit the feature data, with a fitting
precision of about 0.01 dex (with two caveats: the [Fe/M] guess has
about twice this precision because of the relative insensitivity of
the Fe5335 feature to iron, and the [Na/M] value may be falsely
amplified because of interstellar absorption). Model uncertainties
make the accuracies of these values at least twice the magnitude of
the precision.  Forcing scaled-solar abundances does not change the
age very much, but it increases the rms goodness of model-data fit by
a factor of 3 and broadens the allowed range of age to $\pm 1$ Gyr.
The abundance ratios do not show strong trends with radius, except for
the nuclear weakening of CN (measuring mainly N) mentioned above,
which needs reconfirming with better data.  The overall abundance
pattern contrasts with larger elliptical galaxies, in which all
measurable lighter elements are enhanced relative to iron and
calcium. Nucleosynthetic theory does not provide a ready explanation
for these mixtures.

\end{abstract}


\section{Introduction}

M32 is a compact elliptical galaxy of around $10^9 M_\sun$
southeast of the nucleus of the Andromeda galaxy. It shares the
distinction of being the nearest elliptical galaxy with NGC 205,
Andromeda's other elliptical companion. Structurally, M32 lies near an
extrapolation of elliptical galaxy properties, whereas NGC 205, with its
much lower surface brightness, has more in common with disk galaxies
\citep{kor89}. It is usually M32, therefore, that attracts attention
from those interested in the properties, histories, and stellar
populations of elliptical galaxies. Enthusiasm should be tempered by
some caution since M32 is so dense that analogues are rare, and so it
may represent a subclass of elliptical galaxies rather than elliptical
galaxies in general. It probably contains a central black hole of mass
$3 \times 10^6$ M$_\sun$ \citep{ben96}.

A comprehensive understanding of M32's history would connect its
structure and kinematics with its stellar population properties. There
is some dispute in the literature, but most investigations agree that
the nuclear regions of M32 are best fit with a stellar population of
approximately solar composition and an age of, very roughly, 4
Gyr. Published long-slit spectroscopy of M32 (Gonz\'alez 1993,
hereafter G93) does not reach as far as $1\farcm 3$ [1.8 $R_e$(PA);
66\% of light enclosed] from the nucleus where {\em Hubble Space
Telescope} (HST) photometry of resolved red giants was obtained by
\citet{grill96}. Under the assumption of constant age the colors of
the giants map uniquely to abundance, and the abundance spread found
is very similar to that of the solar neighborhood. The best
consistency with {\em extrapolated} G93 spectroscopy was obtained with
old ages of 8 Gyr or older. It should be emphasized that the models
simultaneously matched both the distribution of giant temperatures and
all spectroscopic indices.

Only G93 gives a suitable comparable data set for optical
feature-strength gradients on the Lick/IDS system [system definition
in \citet{wor94,wor97}]. \citet{hardy94} gave long-slit data for M32
but did not transform completely to the Lick/IDS system. \citet{dav91}
measured eight high-quality indices, but his spectra apparently had a
spectral resolution somewhat coarser than Lick/IDS, as evidenced by
the fact that the narrower indices were weaker in comparison
stars. This data could, in principle, be linearly transformed to the
Lick/IDS system, but this has not been attempted here. The
\citet{dav91} results do make a very useful near-differential
comparison set, and we agree with his conclusion that most spectral
indices show gradients in the central 20\arcsec .  \citet{jones94}
report that no index gradients are observed in the central 30\arcsec\ 
but give no further information. High-quality nuclear and near-nuclear
data are given in \citet{trag98}, \citet{delburgo01}, and
\citet{rose94}. This short paper summarizes additional ground-based
(MDM) and space-based (HST) long-slit spectra. The HST {\em Space
Telescope Imaging Spectrograph} (STIS) spectra are discussed in \S 2,
followed by the ground-based data.

\section{Spectra from Space and False Conclusions}

The light profile of M32 (Figure 1) is strongly suggestive of two
components: a compact, dense nuclear component superposed on a
broader, nearly $r^{1/4}$ profile. The inner two arcseconds, about 7
parsecs in radius, almost look like a distinct subgalaxy in the light
profile. It behooves the investigator to look for stellar population
differences at this radius. Kinematic and structural discontinuities
in elliptical galaxies are often marked by changes in the slopes of
absorption feature-strength gradients at the corresponding location
\citep{ben92}.

\begin{figure}
%\epsscale{0.6}
\plotone{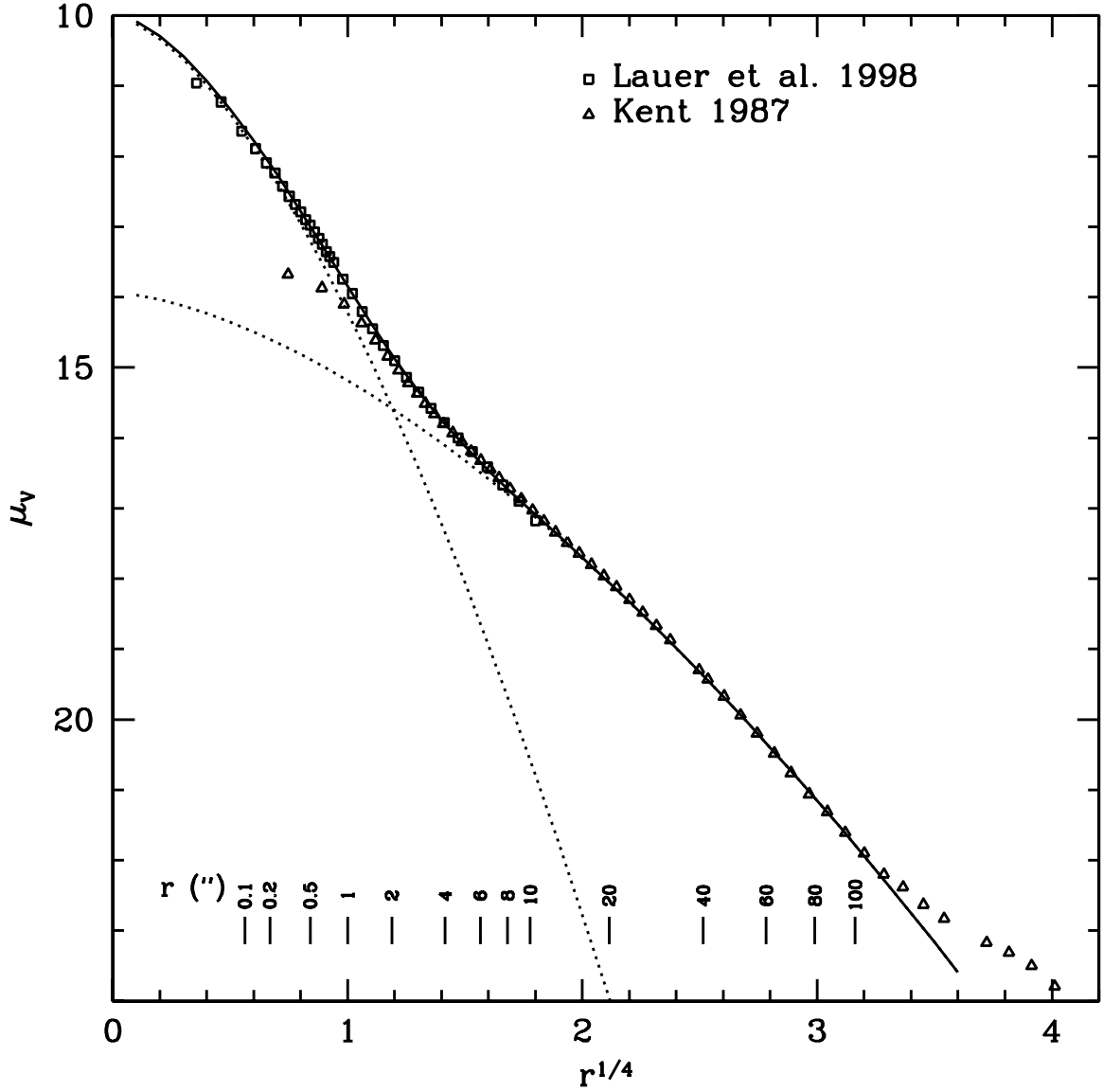}
\caption{ Light profile of M32 from \citet{lauer98} for the
inner portions and \citet{kent87} for the outer portions. A shift of 0.35
mag was applied to the Kent data to transform it to $V$ magnitude. Two
S\'ersic functions ({\em dotted lines}) and their sum ({\em solid line}) are plotted,
indicating a structural discontinuity at around $r=2$\arcsec . The
spatial scale is 3.7 pc arcsecond$^{-1}$.
\label{fig1}}
\end{figure}

\begin{figure}
%\epsscale{0.6}
\plotone{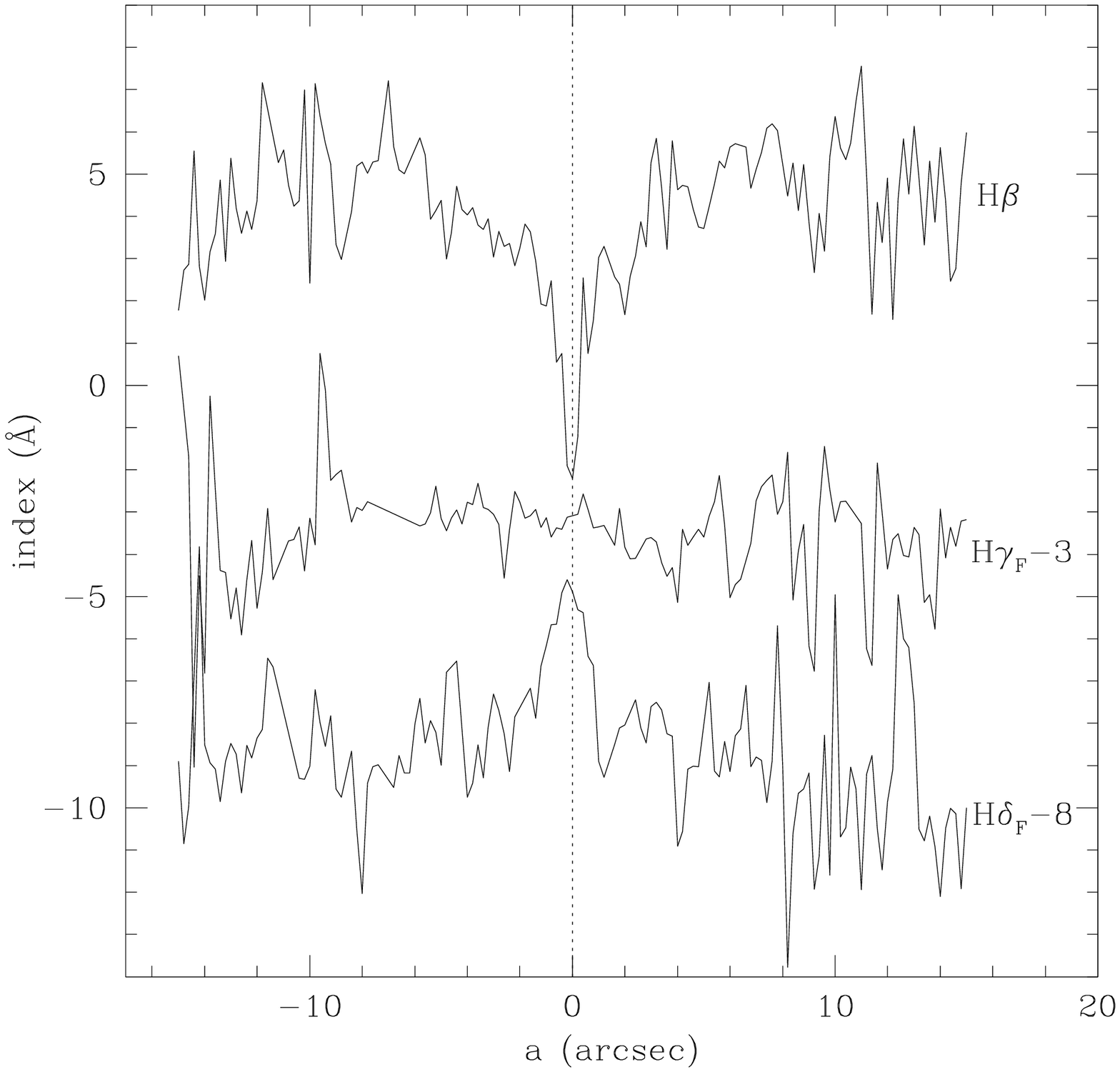}
\caption{ Radial profiles for Lick Balmer indices in M32 as a function
of distance along the slit. The nucleus is at zero. The qualitative
appearance of these three Balmer indices suggests a pattern of
hydrogen emission plus young stars near the nucleus, with H$\beta$
negative (as expected for emission), H$\gamma$ flat, and H$\delta$
positive (as expected for young stars). Appearances are deceiving in
this case, because this pattern is entirely an artifact of residual
cosmic-ray noise in the STIS spectra that just happens, in this case,
to look plausibly astrophysical, at just the scale predicted by
the surface brightness profile.
\label{fig2}}
\end{figure}

The HST STIS data were obtained on 1999 August 6 for program 7438 proposed
by R. W. O'Connell and others [cf. \citet{oc00}]. The bulk of the
exposure time was spent on the ultraviolet portion of the spectrum,
but two exposures (2490 and 3000 s) through a 52$''$ by $0\farcs 2$ slit
were taken with the optical CCD and the G430L disperser. The
dispersion (2.7 \AA pixel$^{-1}$) is perfectly adequate for the measurement
of Lick/IDS indices. The excellent spatial resolution implies many
resolution elements within the $2\arcsec$ central component seen in the
light profile.

Since it turned out to be impossible to measure accurate spectral
indices with this data set, I wish to be brief. After pipeline
calibration and (partial) rejection of cosmic rays through image
combination, spectra were traced, extracted, and wavelength-calibrated
through a cross-correlation technique.  Lick/IDS indices were measured
at 1 pixel intervals along the slit. The result is displayed in
Figure \ref{fig2} for several Balmer indices. The false positive
indication of ionized gas plus young stars in the inner $2\arcsec$
appears to be due entirely to residual cosmic-ray blemishes that were
either too subtle to be removed or appeared on both available
images. The reasons for this conclusion are that (1) other indices
show spikes or dips at different, apparently random radii; (2)
observations symmetric about the center trace are often quite
different; and (3) the large range of data (e.g., $-2$ to 8 \AA\ for
H$\beta$) does not square with ground-based results, in which H$\beta$
lies between 1.9 and 2.3 \AA\ over the whole range of radii.  It
is interesting to note that the total exposure time of the optical
STIS spectra is roughly the same as the sum of the exposures described
below through a telescope of the same aperture, but the errors are at
least an order of magnitude higher.

\section{Spectra from the Ground}
\subsection{Observations}

Ground-based long-slit spectra of M32 suitable for Lick/IDS index
analysis were obtained at the MDM Observatory 2.4 m telescope during
three runs: 1993 September, 1994 October, and 1997 June. The Mark 3
spectrograph was used with blue-sensitive chips (``Charlotte'' in 1993
and 1994 and ``Templeton'' in 1997), with wavelength coverage from
about 3800 to 6100 \AA\ at 2.3 \AA\ pixel$^{-1}$ dispersion. The
resolution was a function of wavelength and run but was always less
than Lick/IDS and therefore easily transformable to the Lick/IDS
system given the nightly set of standard stars observed. Both chips
had the same pixel size, so the spatial scale was constant at $0\farcs
78$ pixel$^{-1}$. The M32 spectra were obtained with the slit oriented
north-south. This is 10$\arcdeg$ away from the major axis of the
galaxy [PA = 170$\arcdeg$; \citet{RC3}] but amounts to a
foreshortening of only 1\%, which is less than the uncertainty in
the spatial scale.

Three exposures in 1993, three in 1994, and four in 1997, with exposure
times ranging from 600 to 1200 s, were bias-subtracted and
flat-fielded using IRAF tasks. Only the 1994 exposures were
sufficiently homogeneous to be co-added. This was done, and we analyzed
the median image in conjunction with the others. The sky was sampled more
than 100$\arcsec$ away and was dominated by terrestrial sky rather than
galaxy light. The galaxy fades by factors of 15$-$40 from the last
extracted spectra at 55$\arcsec$ to where the sky was sampled. Indices
are affected by such self-subtraction in second order. For example, if
the 55$\arcsec$ measurement of H$\beta$ is 2.0 \AA\ but the sky region
of M32 has H$\beta = 1.5$ \AA , the 55$\arcsec$ measurement will be
decreased by, at most, $(2.0 - 1.5)/15 = 0.03$ \AA . If the sky has
the same index value as the target region, no change will result.
Other defects were present in some images. Two images had saturated
pixels near the nucleus, three suffered a sky oversubtraction problem
during processing that had to be corrected by hand, and all images had
a fair number of cosmic ray tracks except for the one median-combined
image from 1994. Seventy 1 pixel wide spectra were extracted
from each side of the nucleus, plus 1 central pixel. These were
cross-correlated with synthetic stellar templates to put them on a
zero-velocity wavelength scale. 

\begin{figure}
%\figurenum{3}
%\epsscale{0.6}
\plottwo{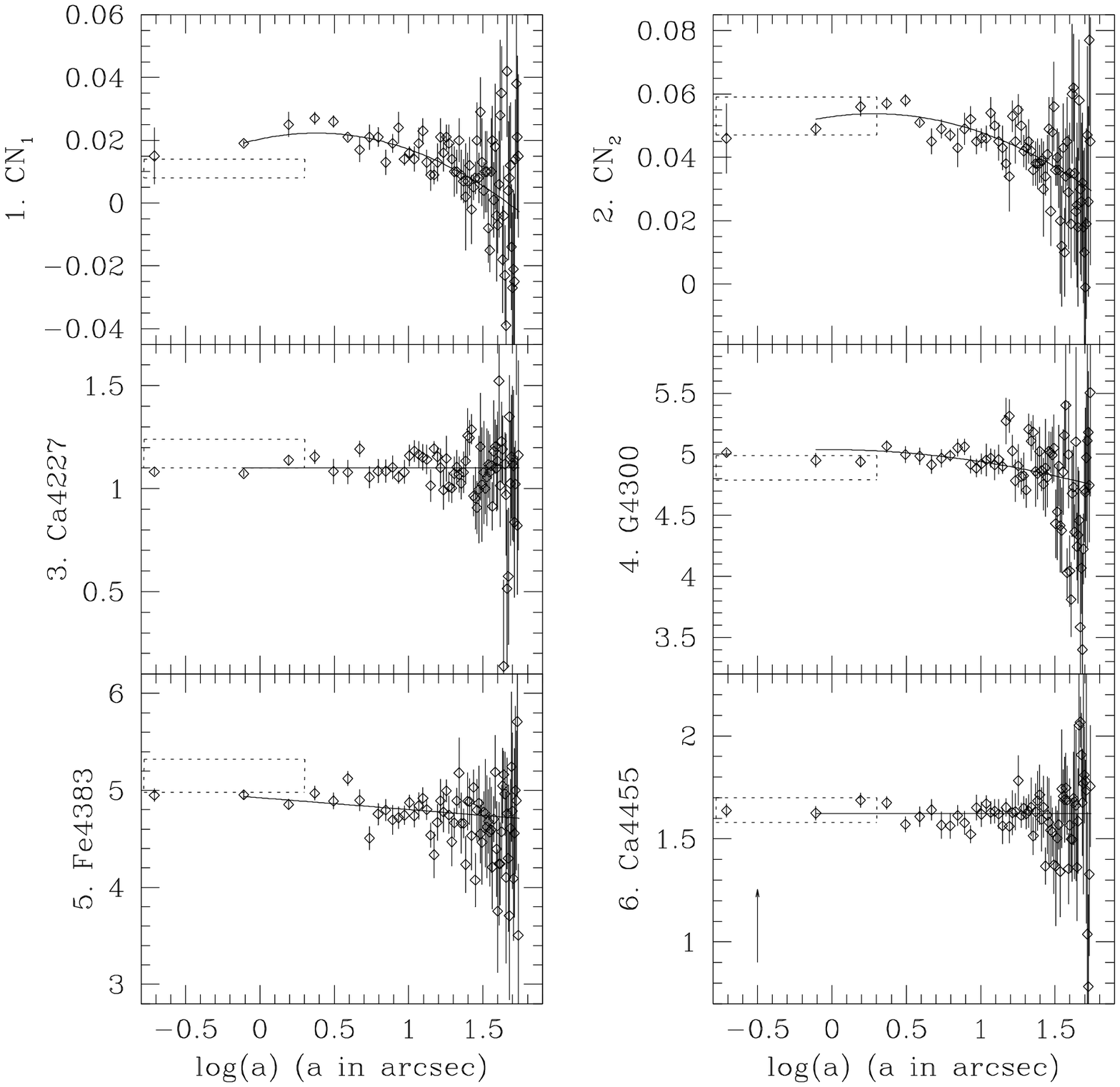}{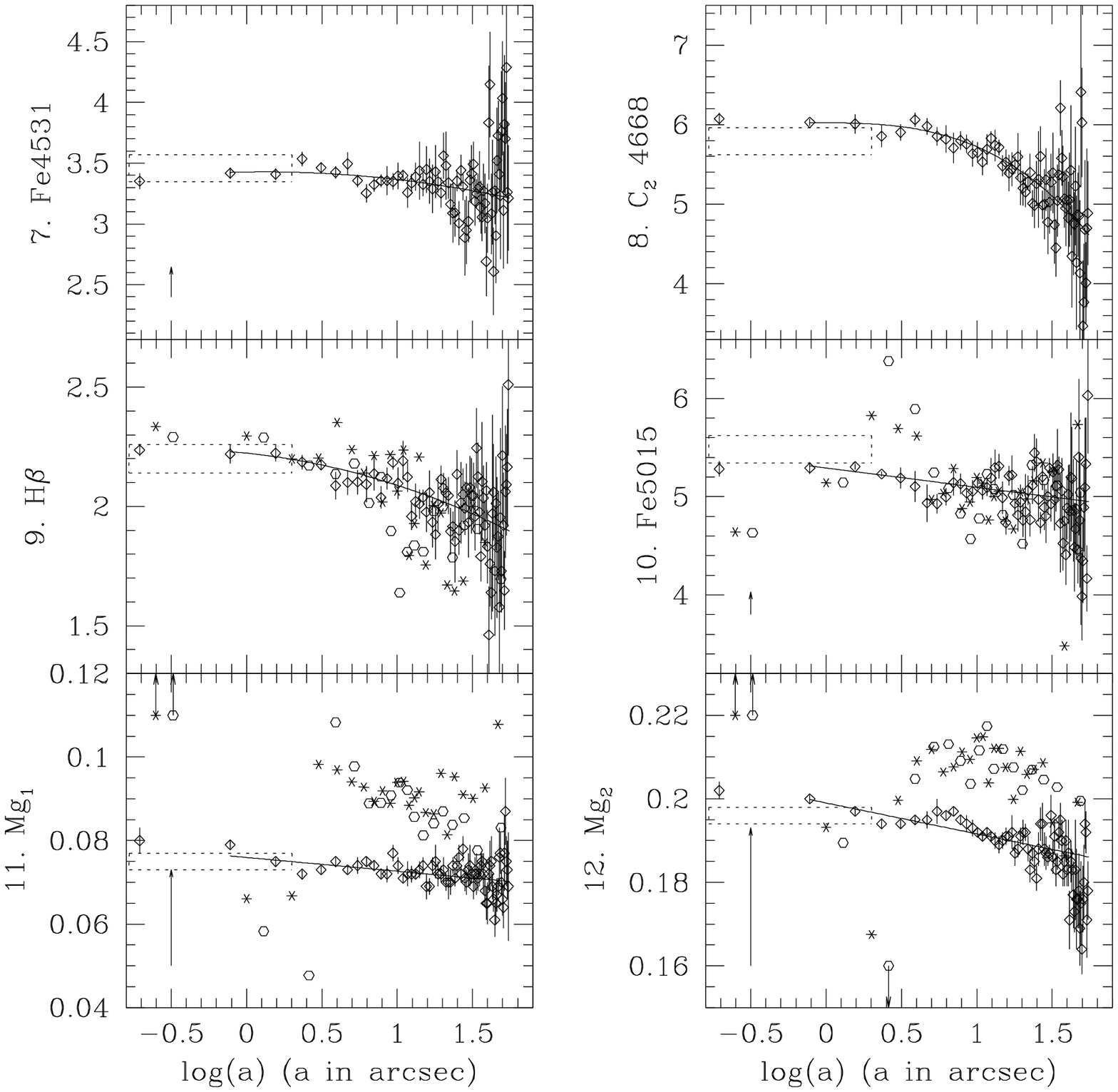}
\caption{Plots of median index values from the MDM data ({\em diamonds}) and their
errors as a function of the logarithm of the semimajor
axis of the isophotal ellipse on which each sampled spectrum lies. The
curve segments are least-square fits to the data.  If there was a
non-negligible correction for systematics, the applied correction is
plotted as a vertical arrow at log$(a)=-0.5$. Nuclear Lick/IDS
measurements from \citet{trag98} (or, in the case of H$\gamma$ and
H$\delta$, unpublished) are indicated as dotted boxes that end at the
nominal Lick/IDS slit length of $4\arcsec/2 = 2\arcsec$ radius. The
boxes enclose $\pm 1 \sigma$. If G93 values exist, they are plotted as
asterisks ({\em major axis}) or open hexagons ({\em minor axis}) with error bars
suppressed. A few G93 data points are indicated as upper or lower
limits if they fall outside the region plotted.
\label{fig3ab}}
\end{figure}

Lick/IDS indices were measured from each spectrum. Each index had 11
independent measurements at each slit location, so with symmetric
north-south pairs analyzed together, all radii except the central
pixel had 22 measurements. The expectation was that most indices would
cluster around the true index value, but some would be affected by a
cosmic ray and would be very wild. The median was therefore adopted as
the statistic of choice. The error was computed by bootstrap
resampling. The complete table of median index values and errors shown
in Fig \ref{fig3ab} is available from the author. Note that the index
TiO$_2$ is not included because it falls past the red end of the MDM
spectra. 

\begin{figure}
\figurenum{3}
%\epsscale{0.6}
\plottwo{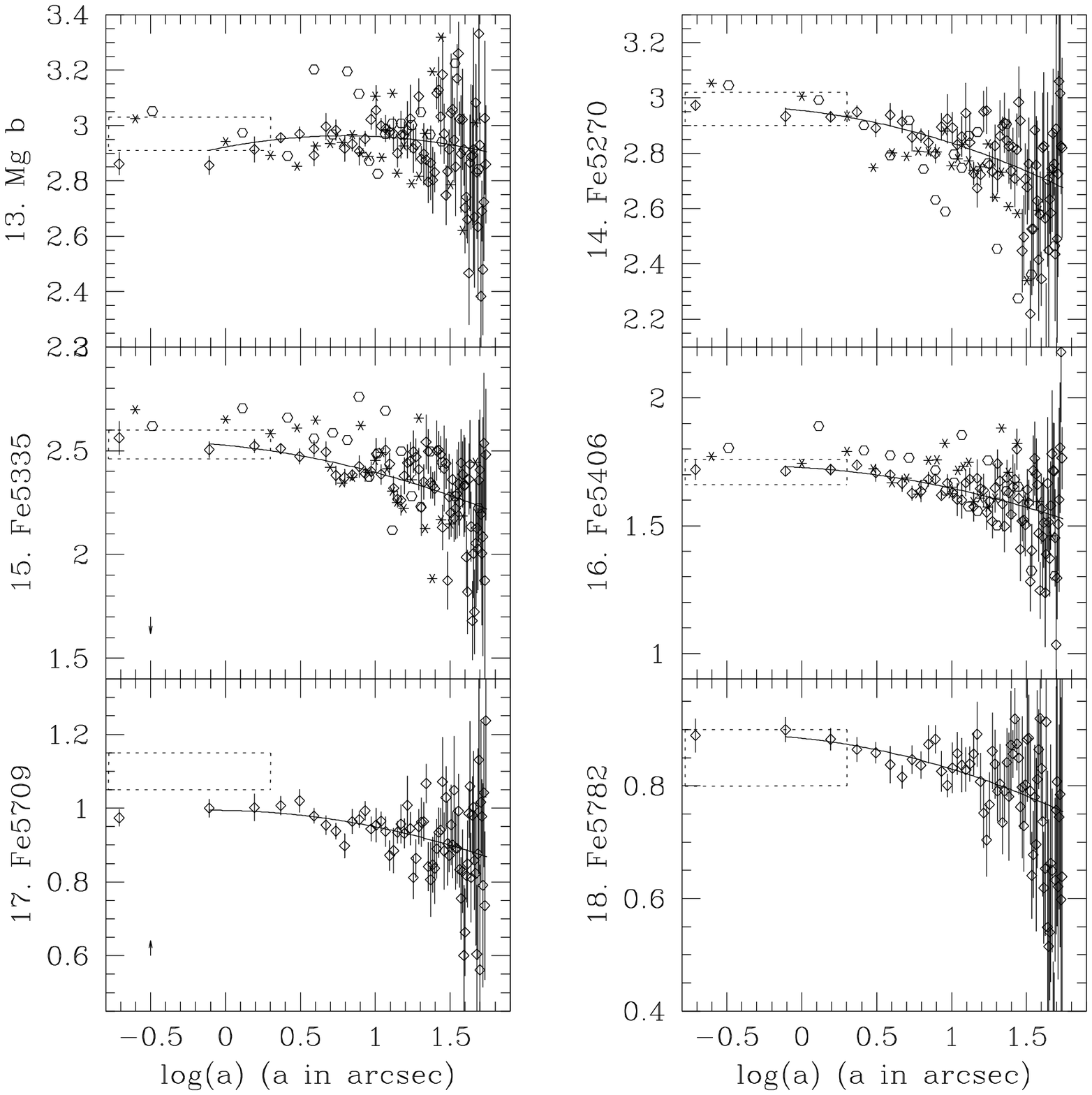}{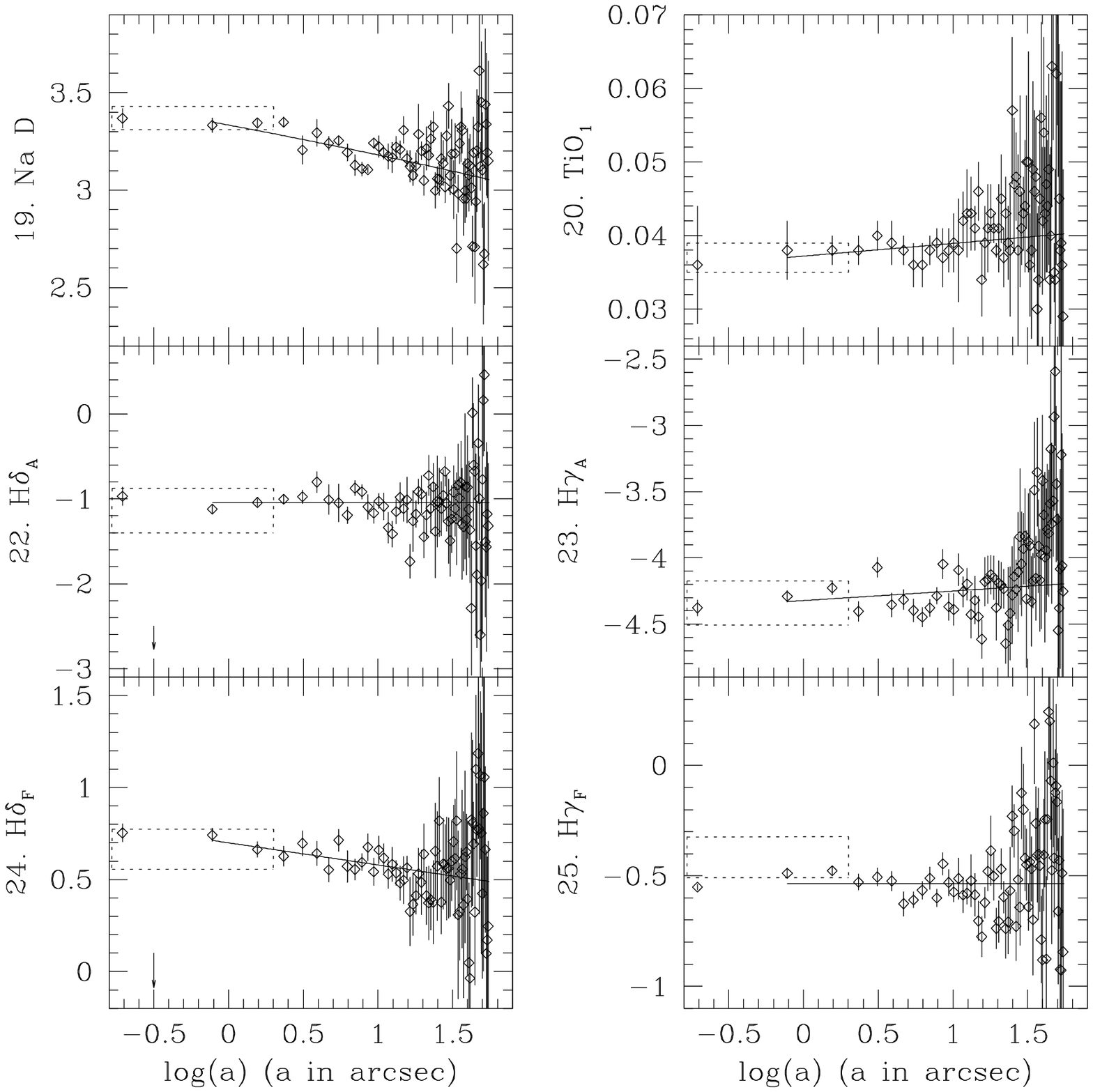}
\caption{ This is a continuation of Figure 3.
\label{fig3cd}}
\end{figure}

Figure \ref{fig3ab} shows the main gradient results, compared with
nuclear values from the original Lick/IDS data set and also G93 for
overlapping indices.  A polynomial-smoothed version of the index
results is plotted as a line and also presented in Table \ref{tab1}.
The errors listed below the index entries in Table \ref{tab1} are
indicative errors for the sum of data points near each tabulated
radius. That is to say, they are larger than the formal error from the
curve fit. It was thought that the data-based error was more
appropriate since many curves could be fit, giving a spread of results
despite formally small errors. Table \ref{tab2} weights the Table
\ref{tab1} results by surface brightness and radius to simulate the
index value measured inside a circular aperture. The last entry in
Table \ref{tab2} is an extrapolation to two effective radii, or about
69\% of light enclosed assuming an $r^{1/4}$ exponential profile.

Comparing indices in common with G93, we note that, as G93 predicted
because of a chromatic focus problem and night-sky lines at Lick
Observatory, Fe5015, Mg$_1$, and Mg$_2$ drift quite a lot. Other
indices match well, except for Mg $b$, discussed immediately
below. Gradient strengths match between the data sets except for
H$\beta$ and Fe5270, for which we obtain shallower slopes.

Fits to all of the data with lines indicate, via the $F$-test, that
Ca4227, Ca4455, H$\delta_A$, and H$\gamma_F$ are statistically
consistent with a value that is constant rather than sloped with
radius. Fits to higher order polynomials indicate, even for these
four, statistically better fits at higher order. However, I decided
against polynomials more complicated than a quadratic to model the
radial trends (except C$_2$4668, for which a cubic was irresistible).
Note that the fact that I find few constant indices is not
inconsistent with \citet{delburgo01}, who detected no radial gradients
in any index, since their data go to a radius of 5\arcsec, equivalent
to the first six points in Figure \ref{fig3ab}, and with larger
errors. Over this spatial range, only the CN indices show a strong
deviation from constant behavior. There is no hint that any stellar
population feature manifests itself at the 2\arcsec\ discontinuity in
the surface brightness profile.

The near-nuclear drop of the CN indices is surprising, given the lack
of any such feature in any other index, and it may or may not be
real. If it is due mostly to a drop in N abundance, then the violet NH
feature gradient data of \citet{dav90} provides support for its
reality since this gradient is also negative. However, whereas the present
data set turns over at 4-5 arcsec, the \citet{dav90} data keep climbing to
the limit of their data at $\approx$11\arcsec. Flattening is possible,
given the error bars, so the two data sets may agree. Alternatively,
future modeling may indicate that age effects dominate over abundance
effects for the NH index, and in that case we should expect it to
weaken toward the nucleus from straightforward age gradient
effects. Clearly, this issue is one that should be resolved with
further observation.

Greater than 1 $\sigma$ zero-point discrepancies occur between this
data set and the original Lick/IDS M32 nuclear data for indices 
Mg $b$ and Fe5709.
Indices Mg$_1$, Mg$_2$, and Na D also were initially
somewhat discrepant, but a second analysis pass modified the
systematic corrections (based on stars and represented by arrows in
Figure \ref{fig3ab}) that were applied. The relatively large
corrections for Mg$_1$ and Mg$_2$ are expected because they are very
broad and subject to the effects of instrumental response more than
other indices. Of the persistently discrepant indices Fe5709 has no
measurements from other sources besides Lick/IDS, so it is hard to
assess which data source might be in error.

The Mg $b$ index, however, may be in some trouble. It is $\sim
2\ \sigma$ weaker than the measurement of G93, $\sim 1.5\ \sigma$ weaker
than Lick/IDS, and $\sim 1\ \sigma$ weaker than
\citet{delburgo01}. However, it is essentially identical to the values
in Figure 9 of \citet{kb99}. In stellar comparisons the Mg $b$ index
was consistent with no systematic drift at all. Additionally, the
sense of the MDM Mg $b$ gradient in the inner few pixels is opposite
that of G93 and also opposite that of Mg$_1$ and Mg$_2$. It is
possible that the MDM data underestimate Mg $b$ by about 0.1 \AA\ in
the inner 2 pixels, or $1\farcs 6$, but not certain.

A large systematic correction was applied to Ca4455 on the basis of stellar
comparisons. This correction was similar for all three data sets and
appears to be a real shift between the original and the MDM
systems. Its origin is unknown.

% Table 1
% abbreviated table of M32 index gradient values

\begin{center}
\begin{deluxetable}{lrrrrrrrrrrrr}
\tabletypesize{\scriptsize}
\tablecolumns{13}
\tablewidth{468pt}
\tablecaption{Polynomial Smoothed Indices and Errors  \label{tab1}}
\tablehead{ \colhead{$a$ (\arcsec) } &
\colhead{CN$_1$} & \colhead{CN$_2$ } & \colhead{Ca4227}  &
\colhead{G4300} & \colhead{Fe4383} & \colhead{Ca4455} &
\colhead{Fe4531} & \colhead{Fe4668} & \colhead{H$\beta$} &
\colhead{Fe5015} & \colhead{Mg$_1$} & \colhead{Mg$_2$} }

\startdata

  0.195 &  0.015 &  0.046 &  1.101 &  5.002 &  5.010 &  1.624 &  3.371 &  6.075 &  2.234 &  5.431 &  0.078 &  0.204 \\
        &  0.009 &  0.012 &  0.015 &  0.021 &  0.062 &  0.032 &  0.067 &  0.076 &  0.026 &  0.074 &  0.003 &  0.005 \\
  0.78  &  0.019 &  0.052 &  1.101 &  5.039 &  4.937 &  1.624 &  3.424 &  6.024 &  2.230 &  5.313 &  0.076 &  0.200 \\
        &  0.001 &  0.002 &  0.014 &  0.057 &  0.049 &  0.035 &  0.043 &  0.041 &  0.040 &  0.058 &  0.001 &  0.001 \\
  1.56  &  0.022 &  0.054 &  1.101 &  5.034 &  4.901 &  1.624 &  3.428 &  6.018 &  2.208 &  5.254 &  0.075 &  0.198 \\
        &  0.004 &  0.003 &  0.025 &  0.042 &  0.044 &  0.038 &  0.047 &  0.125 &  0.034 &  0.029 &  0.001 &  0.001 \\
  2.34  &  0.022 &  0.054 &  1.101 &  5.023 &  4.879 &  1.624 &  3.424 &  5.999 &  2.190 &  5.219 &  0.075 &  0.196 \\
        &  0.002 &  0.002 &  0.032 &  0.050 &  0.073 &  0.032 &  0.060 &  0.142 &  0.030 &  0.028 &  0.001 &  0.001 \\
  4.0   &  0.022 &  0.053 &  1.101 &  5.001 &  4.851 &  1.624 &  3.410 &  5.944 &  2.159 &  5.173 &  0.074 &  0.195 \\
        &  0.001 &  0.002 &  0.035 &  0.052 &  0.066 &  0.035 &  0.054 &  0.066 &  0.038 &  0.126 &  0.001 &  0.001 \\
  6.5   &  0.020 &  0.051 &  1.101 &  4.972 &  4.825 &  1.624 &  3.390 &  5.850 &  2.125 &  5.132 &  0.073 &  0.193 \\
        &  0.002 &  0.002 &  0.027 &  0.032 &  0.063 &  0.031 &  0.030 &  0.073 &  0.031 &  0.052 &  0.001 &  0.001 \\
 10     &  0.017 &  0.048 &  1.101 &  4.940 &  4.803 &  1.624 &  3.366 &  5.722 &  2.089 &  5.095 &  0.073 &  0.192 \\
        &  0.001 &  0.002 &  0.021 &  0.038 &  0.045 &  0.022 &  0.038 &  0.058 &  0.023 &  0.045 &  0.001 &  0.001 \\
 15     &  0.014 &  0.045 &  1.101 &  4.904 &  4.781 &  1.624 &  3.338 &  5.554 &  2.051 &  5.061 &  0.072 &  0.190 \\
        &  0.002 &  0.002 &  0.025 &  0.049 &  0.061 &  0.027 &  0.054 &  0.065 &  0.027 &  0.049 &  0.001 &  0.001 \\
 25     &  0.008 &  0.040 &  1.101 &  4.851 &  4.754 &  1.624 &  3.296 &  5.268 &  1.996 &  5.017 &  0.071 &  0.189 \\
        &  0.002 &  0.002 &  0.034 &  0.054 &  0.072 &  0.035 &  0.049 &  0.081 &  0.029 &  0.050 &  0.001 &  0.001 \\
 40     &  0.002 &  0.034 &  1.101 &  4.794 &  4.730 &  1.624 &  3.250 &  4.917 &  1.940 &  4.977 &  0.071 &  0.187 \\
        &  0.003 &  0.004 &  0.046 &  0.084 &  0.105 &  0.050 &  0.066 &  0.110 &  0.037 &  0.077 &  0.001 &  0.001 \\

\tableline
%\cutinhead{Indices 13 through 20 and 21 through 25}
 $a$ (\arcsec) & Mg $b$ & Fe5270 & Fe5335  & Fe5406 & Fe5709 & Fe5782 &
Na D & TiO$_1$ & H$\delta_A$ & H$\gamma_A$ & H$\delta_F$ & H$\gamma_F$ \\
\tableline

  0.195 &  2.861 &  2.968 &  2.546 &  1.730 &  0.981 &  0.891 &  3.369 &  0.036 & -1.043 & -4.373 &  0.785 & -0.535 \\
        &  0.040 &  0.021 &  0.088 &  0.039 &  0.016 &  0.030 &  0.047 &  0.008 &  0.098 &  0.062 &  0.044 &  0.015 \\
  0.78  &  2.910 &  2.960 &  2.533 &  1.731 &  0.994 &  0.887 &  3.349 &  0.037 & -1.043 & -4.329 &  0.713 & -0.535 \\
        &  0.036 &  0.037 &  0.041 &  0.020 &  0.025 &  0.022 &  0.037 &  0.003 &  0.057 &  0.045 &  0.036 &  0.025 \\
  1.56  &  2.939 &  2.940 &  2.510 &  1.719 &  0.991 &  0.878 &  3.306 &  0.038 & -1.043 & -4.308 &  0.676 & -0.535 \\
        &  0.048 &  0.020 &  0.032 &  0.021 &  0.038 &  0.021 &  0.025 &  0.002 &  0.061 &  0.051 &  0.045 &  0.021 \\
  2.34  &  2.951 &  2.924 &  2.492 &  1.709 &  0.986 &  0.871 &  3.279 &  0.038 & -1.043 & -4.295 &  0.655 & -0.535 \\
        &  0.017 &  0.023 &  0.025 &  0.013 &  0.027 &  0.022 &  0.027 &  0.002 &  0.054 &  0.075 &  0.050 &  0.037 \\
  4.0   &  2.960 &  2.897 &  2.462 &  1.690 &  0.976 &  0.858 &  3.244 &  0.038 & -1.043 & -4.278 &  0.627 & -0.535 \\
        &  0.031 &  0.030 &  0.032 &  0.018 &  0.017 &  0.019 &  0.039 &  0.002 &  0.108 &  0.059 &  0.047 &  0.035 \\
  6.5   &  2.963 &  2.867 &  2.429 &  1.669 &  0.963 &  0.844 &  3.211 &  0.039 & -1.043 & -4.263 &  0.602 & -0.535 \\
        &  0.020 &  0.020 &  0.020 &  0.015 &  0.014 &  0.014 &  0.020 &  0.001 &  0.064 &  0.038 &  0.033 &  0.022 \\
 10     &  2.960 &  2.837 &  2.395 &  1.647 &  0.950 &  0.830 &  3.180 &  0.039 & -1.043 & -4.250 &  0.579 & -0.535 \\
        &  0.017 &  0.030 &  0.022 &  0.018 &  0.015 &  0.016 &  0.019 &  0.002 &  0.060 &  0.047 &  0.032 &  0.026 \\
 15     &  2.954 &  2.804 &  2.360 &  1.623 &  0.934 &  0.815 &  3.152 &  0.039 & -1.043 & -4.237 &  0.558 & -0.535 \\
        &  0.026 &  0.024 &  0.028 &  0.021 &  0.017 &  0.018 &  0.023 &  0.001 &  0.068 &  0.058 &  0.041 &  0.034 \\
 25     &  2.940 &  2.758 &  2.309 &  1.589 &  0.911 &  0.793 &  3.114 &  0.040 & -1.043 & -4.221 &  0.532 & -0.535 \\
        &  0.026 &  0.028 &  0.031 &  0.024 &  0.021 &  0.018 &  0.026 &  0.002 &  0.091 &  0.083 &  0.058 &  0.039 \\
 40     &  2.921 &  2.711 &  2.258 &  1.554 &  0.887 &  0.771 &  3.079 &  0.040 & -1.043 & -4.206 &  0.507 & -0.535 \\
        &  0.036 &  0.042 &  0.045 &  0.032 &  0.028 &  0.029 &  0.040 &  0.003 &  0.131 &  0.100 &  0.077 &  0.054 \\

\enddata
\end{deluxetable}
\end{center}

%----------------------------------------------------------------------
% Table 2
% table of M32 synthetic aperture index values

\begin{center}
\begin{deluxetable}{lrrrrrrrrrrrr}
\tabletypesize{\scriptsize}
\tablecolumns{13}
\tablewidth{468pt}
\tablecaption{Synthetic Circular Aperture Indices and Errors  \label{tab2}}
\tablehead{ \colhead{$R/R_e$ } &
\colhead{CN$_1$} & \colhead{CN$_2$ } & \colhead{Ca4227}  &
\colhead{G4300} & \colhead{Fe4383} & \colhead{Ca4455} &
\colhead{Fe4531} & \colhead{Fe4668} & \colhead{H$\beta$} &
\colhead{Fe5015} & \colhead{Mg$_1$} & \colhead{Mg$_2$} }

\startdata

  0.1 &  0.021 &  0.053 &  1.101 &  5.021 &  4.902 &  1.624 &  3.416 &  6.002 &  2.199 &  5.256 &  0.075 &  0.198 \\
      &  0.002 &  0.003 &  0.017 &  0.033 &  0.041 &  0.024 &  0.037 &  0.060 &  0.024 &  0.045 &  0.001 &  0.001 \\
 0.125&  0.021 &  0.052 &  1.101 &  5.016 &  4.892 &  1.624 &  3.414 &  5.987 &  2.191 &  5.240 &  0.075 &  0.197 \\
      &  0.002 &  0.002 &  0.019 &  0.035 &  0.045 &  0.026 &  0.038 &  0.062 &  0.026 &  0.053 &  0.001 &  0.001 \\
  0.25&  0.020 &  0.051 &  1.101 &  4.991 &  4.860 &  1.624 &  3.399 &  5.904 &  2.156 &  5.188 &  0.074 &  0.195 \\
      &  0.002 &  0.002 &  0.021 &  0.035 &  0.049 &  0.026 &  0.037 &  0.063 &  0.027 &  0.052 &  0.001 &  0.001 \\
  0.5 &  0.017 &  0.048 &  1.101 &  4.953 &  4.827 &  1.624 &  3.372 &  5.747 &  2.111 &  5.135 &  0.073 &  0.193 \\
      &  0.002 &  0.002 &  0.023 &  0.041 &  0.054 &  0.027 &  0.042 &  0.065 &  0.027 &  0.051 &  0.001 &  0.001 \\
  1.0 &  0.013 &  0.045 &  1.101 &  4.909 &  4.798 &  1.624 &  3.339 &  5.526 &  2.063 &  5.088 &  0.072 &  0.191 \\
      &  0.002 &  0.002 &  0.029 &  0.050 &  0.065 &  0.032 &  0.048 &  0.075 &  0.029 &  0.055 &  0.001 &  0.001 \\
  2.0 &  0.008 &  0.040 &  1.101 &  4.860 &  4.772 &  1.624 &  3.300 &  5.246 &  2.012 &  5.046 &  0.072 &  0.190 \\
      &  0.003 &  0.004 &  0.038 &  0.070 &  0.088 &  0.042 &  0.058 &  0.095 &  0.034 &  0.071 &  0.001 &  0.001 \\

\tableline
 $R/ Re$  & Mg $b$ & Fe5270 & Fe5335  & Fe5406 & Fe5709 & Fe5782 &
Na D & TiO$_1$ & H$\delta_A$ & H$\gamma_A$ & H$\delta_F$ & H$\gamma_F$ \\
\tableline

  0.1 &  2.933 &  2.933 &  2.503 &  1.713 &  0.986 &  0.875 &  3.300 &  0.038 & -1.043 & -4.309 &  0.678 & -0.535 \\
      &  0.024 &  0.019 &  0.029 &  0.015 &  0.017 &  0.016 &  0.024 &  0.002 &  0.052 &  0.040 &  0.031 &  0.019 \\
 0.125&  2.937 &  2.925 &  2.494 &  1.708 &  0.983 &  0.871 &  3.289 &  0.038 & -1.043 & -4.303 &  0.668 & -0.535 \\
      &  0.024 &  0.020 &  0.028 &  0.015 &  0.017 &  0.016 &  0.025 &  0.002 &  0.058 &  0.042 &  0.032 &  0.021 \\
  0.25&  2.947 &  2.895 &  2.460 &  1.688 &  0.972 &  0.857 &  3.251 &  0.038 & -1.043 & -4.284 &  0.636 & -0.535 \\
      &  0.022 &  0.022 &  0.026 &  0.016 &  0.016 &  0.016 &  0.023 &  0.002 &  0.060 &  0.042 &  0.033 &  0.022 \\
  0.5 &  2.949 &  2.856 &  2.417 &  1.660 &  0.956 &  0.839 &  3.209 &  0.039 & -1.043 & -4.264 &  0.603 & -0.535 \\
      &  0.023 &  0.024 &  0.026 &  0.018 &  0.017 &  0.016 &  0.023 &  0.002 &  0.065 &  0.050 &  0.037 &  0.027 \\
  1.0 &  2.943 &  2.815 &  2.372 &  1.630 &  0.937 &  0.820 &  3.171 &  0.039 & -1.043 & -4.247 &  0.575 & -0.535 \\
      &  0.026 &  0.028 &  0.030 &  0.021 &  0.019 &  0.019 &  0.027 &  0.002 &  0.081 &  0.064 &  0.047 &  0.034 \\
  2.0 &  2.930 &  2.772 &  2.325 &  1.598 &  0.915 &  0.800 &  3.135 &  0.039 & -1.043 & -4.231 &  0.549 & -0.535 \\
      &  0.032 &  0.036 &  0.039 &  0.027 &  0.024 &  0.025 &  0.035 &  0.003 &  0.108 &  0.079 &  0.062 &  0.044 \\

\enddata
\end{deluxetable}
\end{center}

%--------------------------------------------------------------------

\subsection{Mean Age}

\begin{figure}
%\epsscale{0.6}
\plotone{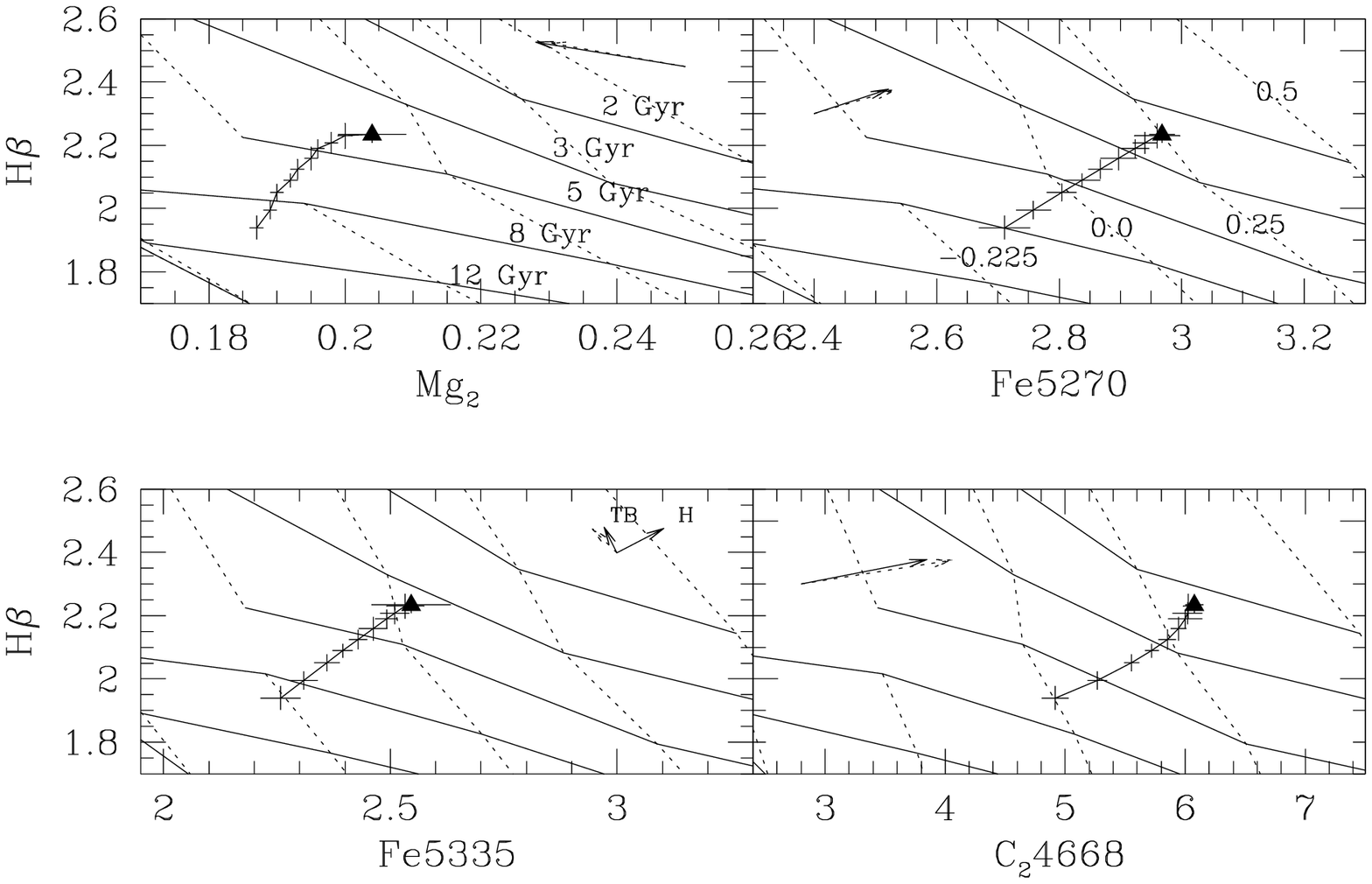}
\caption{Plots of four age-metal diagnostic diagrams that array
H$\beta$ against metal features. M32 spectral data from Table
\ref{tab1} are plotted, with the central pixel marked as a
triangle and subsequent samples marked only by error bars.
\citet{w94} models are plotted as a grid with
same-age lines plotted with solid lines and same-abundance lines
plotted with dotted lines. The ages are marked in the Mg$_2$ panel, and
the [M/H] values are marked in the Fe5270 panel. The vectors drawn with
dotted lines represent the vector between the model for (age, [M/H]) =
(4.75 Gyr, 0.02) and the table 2 nuclear-aperture observation. The
vectors drawn with solid lines represent a model shift caused by
the slightly nonsolar abundance pattern that is discussed in the
next section. The ``TB'' and ``H'' choices in the Fe5335 panel stand
for predictions based on TB95 and \citet{houda} synthetic
stellar fluxes. The \citet{houda} fluxes were used for all other panels. 
\label{fig4}}
\end{figure}

The basic technique of arriving at a mean-age, mean-abundance pair
using Lick/IDS indices was articulated by \citet{w94} and expanded on
by many authors, among them \citet{delburgo01}, \citet{trag00},
\citet{tant98}, and \citet{thom03}. An unofficial median of ages
quoted for the stellar population in the nucleus of M32, including
other methods of spectral comparison (e.g. \citet{oc80},
\citet{rose94}, \citet{bica90}, and \citet{vaz99}), is around 4
Gyr. Quoting a mean age does not, of course, mean that M32 sprang into
existence 4 Gyr ago. Galaxies are complex systems with considerable
chemodynamic history behind them and we should expect abundance
distributions (probably fairly regular) and age distributions
(probably irregular) for them. The mean age and abundance are
nevertheless convenient signposts, at least for statistical studies.

Effects that would strengthen Balmer-line strengths by adding warm or
hot stars, such as unexpected horizontal-branch morphology or blue
straggler stars, are conveniently muted in M32 compared with large,
redder elliptical galaxies: regarding extreme horizontal-branch star
contributions, \citet{b88} find that M32 has the least mid-UV flux of
any galaxy in their sample, and \citet{rose94} finds that an
intermediate-age population plus a frosting of metal-poor stars
accounts for all of the observables available to him at high spectral
resolution.  Furthermore, G93 found negligible nebular emission in
M32. Again, M32 seems safe because its Balmer lines are already
strong, so any possible emission line spectrum would have to be fairly
powerful to affect our interpretation. Finally, nonsolar abundance
ratio effects are mild in M32 (see below). This greatly reduces the
danger of significant modulations of either the main-sequence turnoff
temperature or the giant branch temperature, the two quantities
fundamentally measured by the Balmer-metal technique, as a result of
altered chemical mixtures causing changes in the structures of the
stars.

Some Balmer-metal diagrams are shown in Figure \ref{fig4}, using
H$\beta$ as an example. Use of the other four Balmer indices gives
virtually identical results. What Figure \ref{fig4} shows is that
(1) inferred age depends on which metal feature is
used. Mg$_2$ and Fe5335 give ages of around 4 Gyr whereas Fe5270 and
C$_2$4668 yield ages closer to 2.5 Gyr. (2) Inferred abundance also
shifts. (3) Age increases and abundance drops with radius. In this
case the changes are very similar from diagram to diagram, with the
abundance always ranging somewhat more than one 0.25 dex interval, and
the age spanning factors of 2.5$-$3 from the nucleus to the last
measured point at 1 $R_e$. 

The shifts from panel to panel in Figure \ref{fig4} are due to
abundance ratio changes, discussed more fully in the next section. The
younger ages in the Fe5270 and C$_2$4668 panels are due to increased
index strength due to enhanced carbon, so the older ages are the more
reliable in this case. This is tempered by uncertainty in oxygen
abundance, which we do not yet measure, but which can effect changes
in isochrone shape [cf. \citet{w98}]. Forcing adherence to
scaled-solar index models, 4 Gyr would be our best guess with the current
data and models for the nucleus, with a mean abundance just slightly
more than solar, say +0.05. This quickly fades with radius to 8-10 Gyr
age and abundance of -0.25 for a radius of 1 $R_e$ = $44\farcs 4$ along
the major axis, or 164 pc from the nucleus. The corresponding minor
axis distance would be $34\farcs 2$, or 127 parsec.

The age and metal abundance hover around those discussed in
\citet{grill96} for their HST field at 1.8 $R_e$. The Lick indices
extrapolated from G93 data indicated age and abundance of (8.5 Gyr,
$-0.25$). Our data would extrapolate to older and slightly more
metal-poor, about (10-11 Gyr, $-0.35$) using the [MgFe] composite
index. From the \citet{grill96} photometry, and assuming 10 Gyr
isochrones, the $V-I_C$ colors of the giants indicate a peak abundance
of $-0.2$. Realizing that the metal-poor side of this peak will
contribute somewhat more than the metal-rich side to the spectroscopic
mean age via greater luminosity and greater contribution to Balmer
strengths, the two approaches are probably still fairly closely in
harmony with each other, although a detailed revisit to the problem
would be a good idea. The older extrapolated age from the new
spectroscopy is something of a surprise since the new H$\beta$
gradient is shallower than G93, but other small changes, such as a
shallower Fe5270 gradient and a depressed Mg $b$ index, compensate for
this.

\subsection{Abundance Ratios}

A discussion of abundance ratio patterns in early-type
galaxies is found in \citet{w98}. The major trend is that larger
elliptical galaxies have deviations from scaled-solar ratios in the
sense that light elements are enhanced relative to Fe-peak
elements. Since not every light element can presently be measured,
this is partly an assumption based on yield predictions from Type Ia
and Type II supernovae. Low-mass elliptical galaxies, like M32, have nearly
solar element ratios.

Diagrams that plot metal index against metal index are usually highly
degenerate with respect to age and scaled-abundance changes. This
leaves such diagrams sensitive to changes that are not simply
different combinations of age and metallicity, namely the modulation
of abundance ratios, such as [Mg/Fe] [cf. the ``age-metallicity
degeneracy'' and ``theorem of sensitivity'' of \citet{w94}]. Models
that track many abundance mixtures with perfect self-consistency
(meaning that all ingredients from opacities to stellar isochrones to
stellar fluxes would be produced with the same element mixtures) have
not yet been produced, but brave forays in this direction have been
made \citep{trag00,tmb03}. For this paper we stay with the
scaled-solar models of \citet{w94}, augmented by the calculations
of nonsolar-mixture synthetic stellar calculations of \citet{houda},
which are very similar to those of \citet{tb95} (hereafter TB95).

An issue uncovered with the TB95 synthetic fluxes should be
mentioned here. Their carbon enhancement was 0.3 dex, sufficient to
make the number abundance of carbon exceed that of oxygen. The
large dissociation energy of the CO molecule causes most of the carbon
to be locked up in this molecule in oxygen-rich stars, but in
carbon-rich stars C$_2$ Swan bands begin to dominate the visible
spectrum in a nonlinear way, approximately as the square of the carbon
abundance. The responses for carbon in TB95's tables 4, 5, and 6 are
therefore significantly overestimated. We double checked this
conclusion with sets of synthetic spectra kindly provided by A. Korn, by
M. Briley, and also by E. Baron. The \citet{houda} carbon responses
were computed with a carbon enhancement of 0.15 dex and thus do not
change the molecular equilibrium very much.

In galaxy data like those considered here, abundance ratios can be
estimated from scaled-solar models by assuming that any deviation
between the model and the observations results from an individual
element enhancement that acts like an overall abundance increase, as
seen, for instance, in \citet{wfg92}. This works only roughly, as (1)
the Lick/IDS indices cover many blended lines and therefore other
species besides the dominant one contribute and (2) abundance mixture
changes the underlying isochrone. \citet{trag00} and \citet{tmb03}
effectively calibrate against line blends by using the results of
TB95, who explored the effects of element ratio changes in the
spectra of three representative stars. The approach taken here is
essentially the same as \citet{trag00} and \citet{tmb03}, except that
we use \citet{houda} spectra.

Using the rms fit of model versus data (the Table \ref{tab2} nuclear
data point) as a figure of merit, a best-fit age, overall
``metallicity'' [M/H], and abundance mixture were found. The best age
is 4.75 Gyr, with [M/H]=+0.02. Small abundance changes were sought
simultaneously to improve the fit. The primary ones are [C/M]=+0.077,
[N/M]=$-0.13$, [Mg/M]=$-0.18$, and [Na/M]=+0.12. These are relatively
well measured, since they affect a variety of indices in a substantial
way. The sodium abundance, of course, could be spurious because of
interstellar absorption in the Na D feature, but the other
measurements are good to roughly 0.02 dex, for which most of the
uncertainty is in the models, not the data. Elements with much larger
uncertainty because of the fact that they do not strongly impact the
Lick/IDS indices are [O/M]=[Fe/M]=[Ca/M]=[Si/M]$\approx 0$,
[Cr/M]$\approx -0.15$, and [Ti/M]$\approx -0.2$. The indices that were
used to find this solution (CN$_2$, Ca4227, Fe4383, C$_2$4668, Mg$_2$,
Fe5270, Fe5335, Fe5406, Fe5709, Na D, and all five Balmer indices) had a
final rms fit of 0.26 in units of the \citet{wor94} standard Lick/IDS
errors. Including all indices raises the rms fit to 0.43, in which the
worst-fitting ones were Mg$_1$ and TiO$_1$. Both of these indices are
sensitive to M giant numbers and temperatures and therefore suffer
increased model uncertainty. This gives a preliminary and rough guess
as to the ultimate accuracy of the \citet{w94} models of better than half of a
Lick/IDS $\sigma$. This is extremely encouraging. However, expressed
in units of the accuracy of the present set of M32 {\em data}, the rms
of 0.43 becomes 3.8. That is, the data far outstrip today's models in
potential accuracy.

\begin{figure}
\plotone{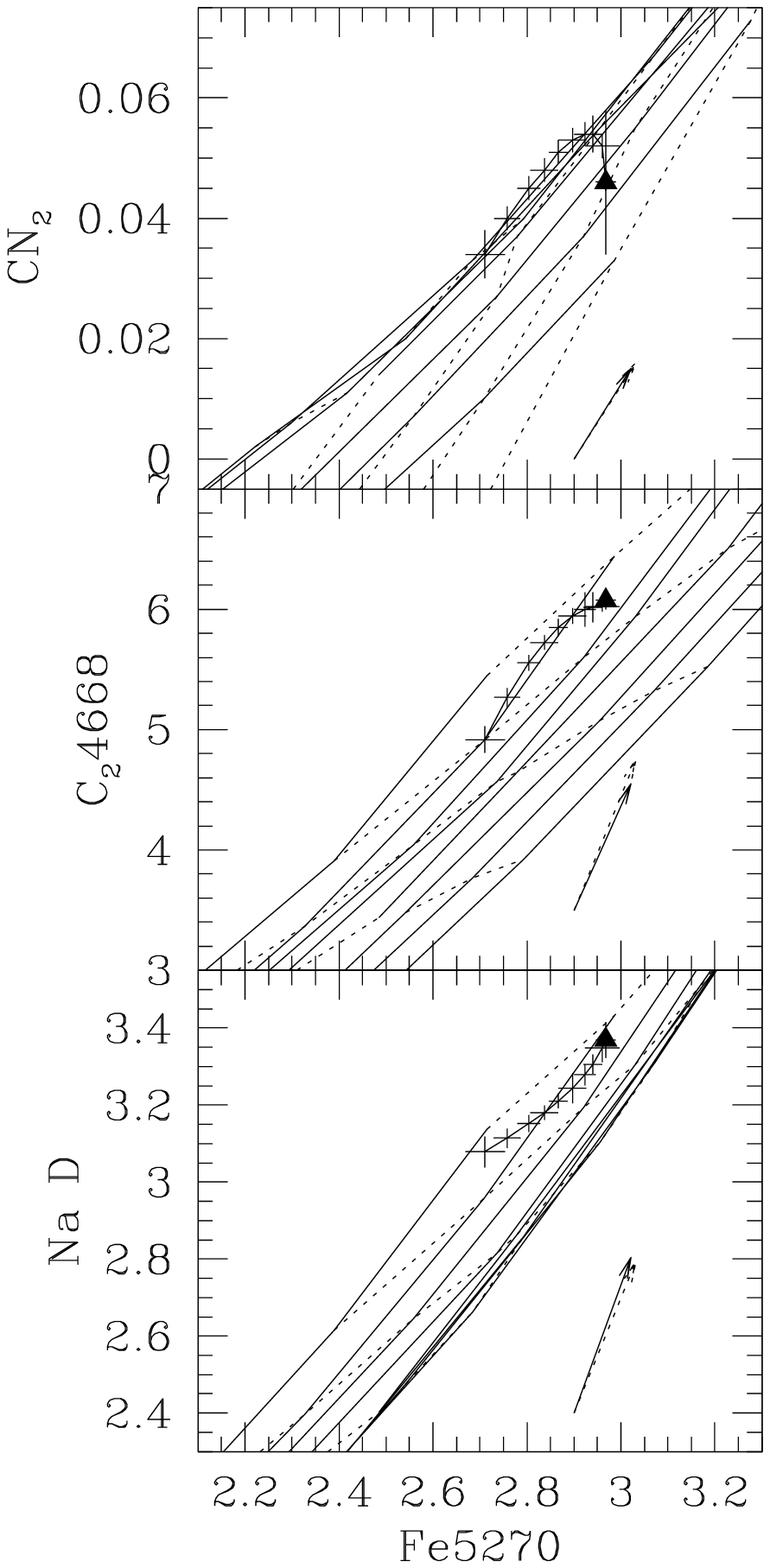}
\caption{ Index-index plots of abundance ratio effects in
M32. Symbols are as in Figure \ref{fig4}: {\em triangle}, nucleus;
{\em error bars}, other samples; {\em grid}, models.
Model ages ({\em solid lines}) are 1, 1.5, 2, 3, 5, 8, 12, and 17 Gyr,
and the top four [M/H] values ({\em dotted lines}) are $-0.25$, 0, 0.25 and
0.5. Increasing age or metallicity increases metallic feature
strengths. Dotted vectors are the offset between an (age, [M/H]) = (4.75
Gyr, 0.02) model and the Table \ref{tab2} nuclear aperture
observation. Solid vectors are the shifts predicted by considering a
nonsolar pattern of [C/M]=+0.077, [N/M]=$-$0.13, [Mg/M]=$-0.18$, and
[Na/M]=+0.12, as discussed in the text. 
\label{fig5}}
\end{figure}

\begin{figure}
\plotone{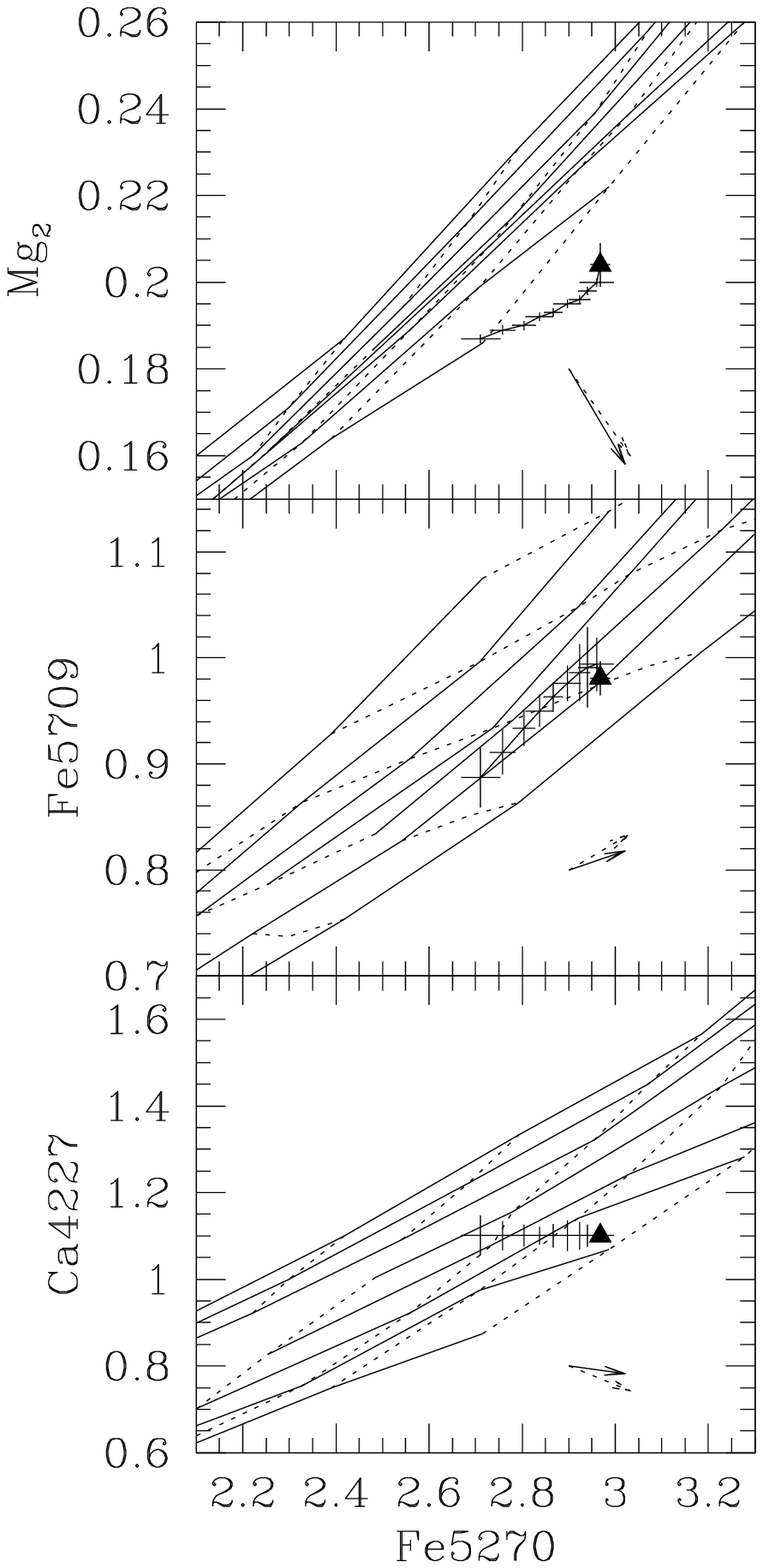}
\caption{ Index-index plots, with models and gradient data. Symbols
are as in Figure \ref{fig5}. 
\label{fig6}}
\end{figure}

\begin{figure}
\plotone{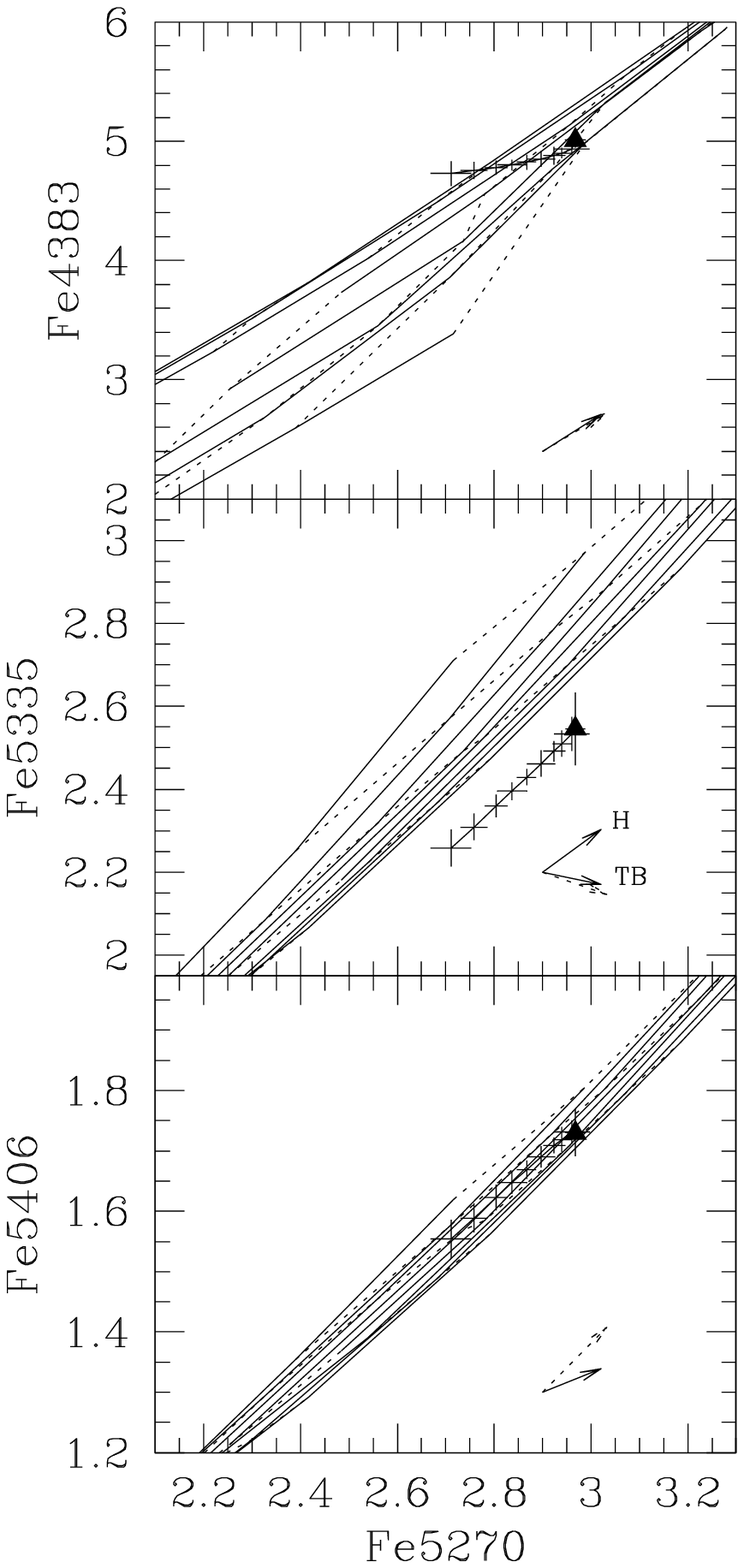}
\caption{ Index-index plots of abundance ratio effects. Symbols
are as in Figure \ref{fig5}. The observations should overlay the model
grid in these Fe-index vs. Fe-index diagrams if Fe were the sole
driver in the feature strengths. The two solid vectors in the Fe5335
panel show different predictions from the synthetic spectra
of TB95 vs.  \citet{houda}.  
\label{fig7}}
\end{figure}

\begin{figure}
\plotone{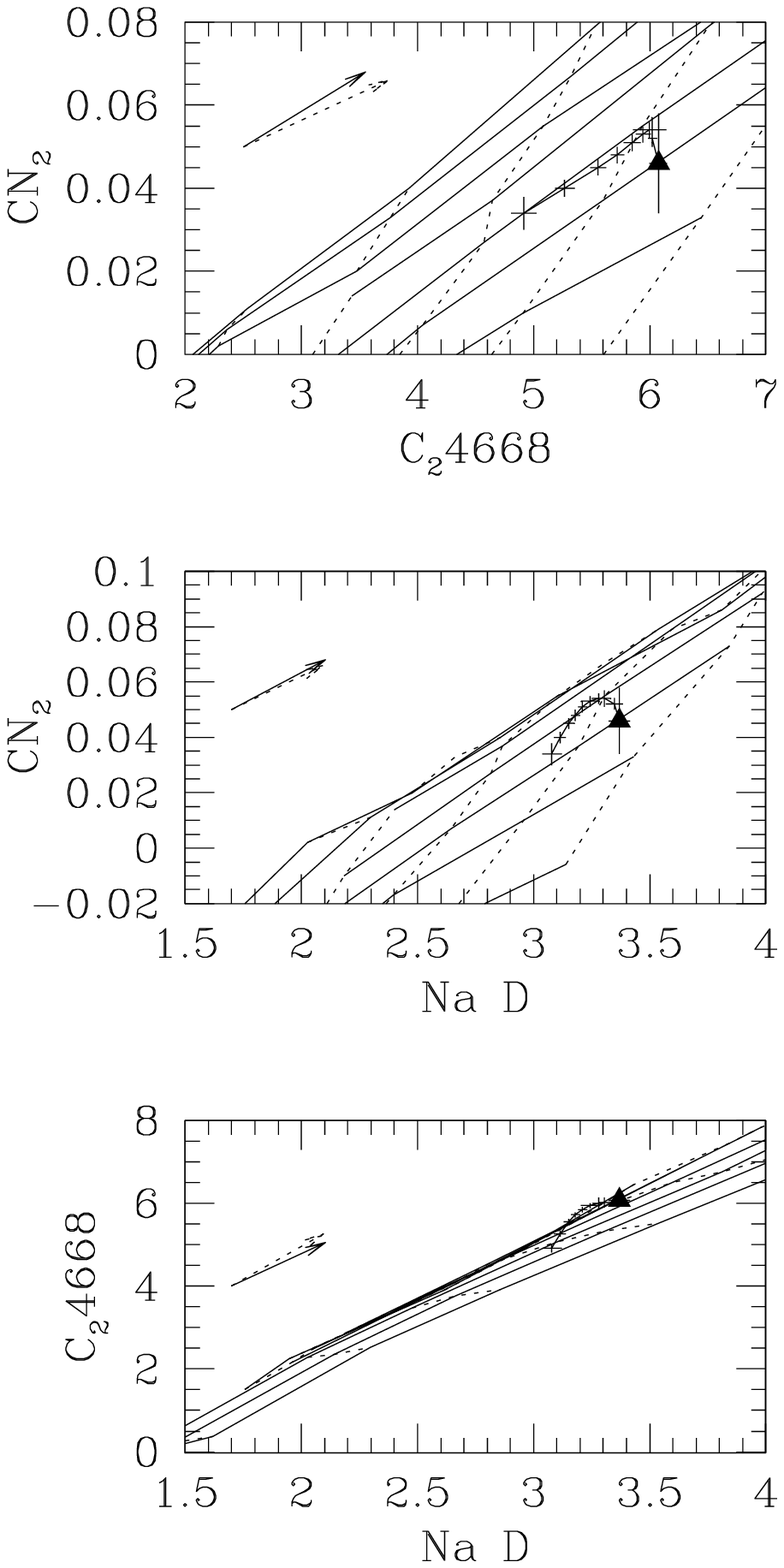}
\caption{ Index-index plots of the interplay between
elements N, C, and Na. The CN index is affected by both the C and N abundances.
Symbols are as in Figure \ref{fig5}.
\label{fig8}}
\end{figure}

\begin{figure}
\plotone{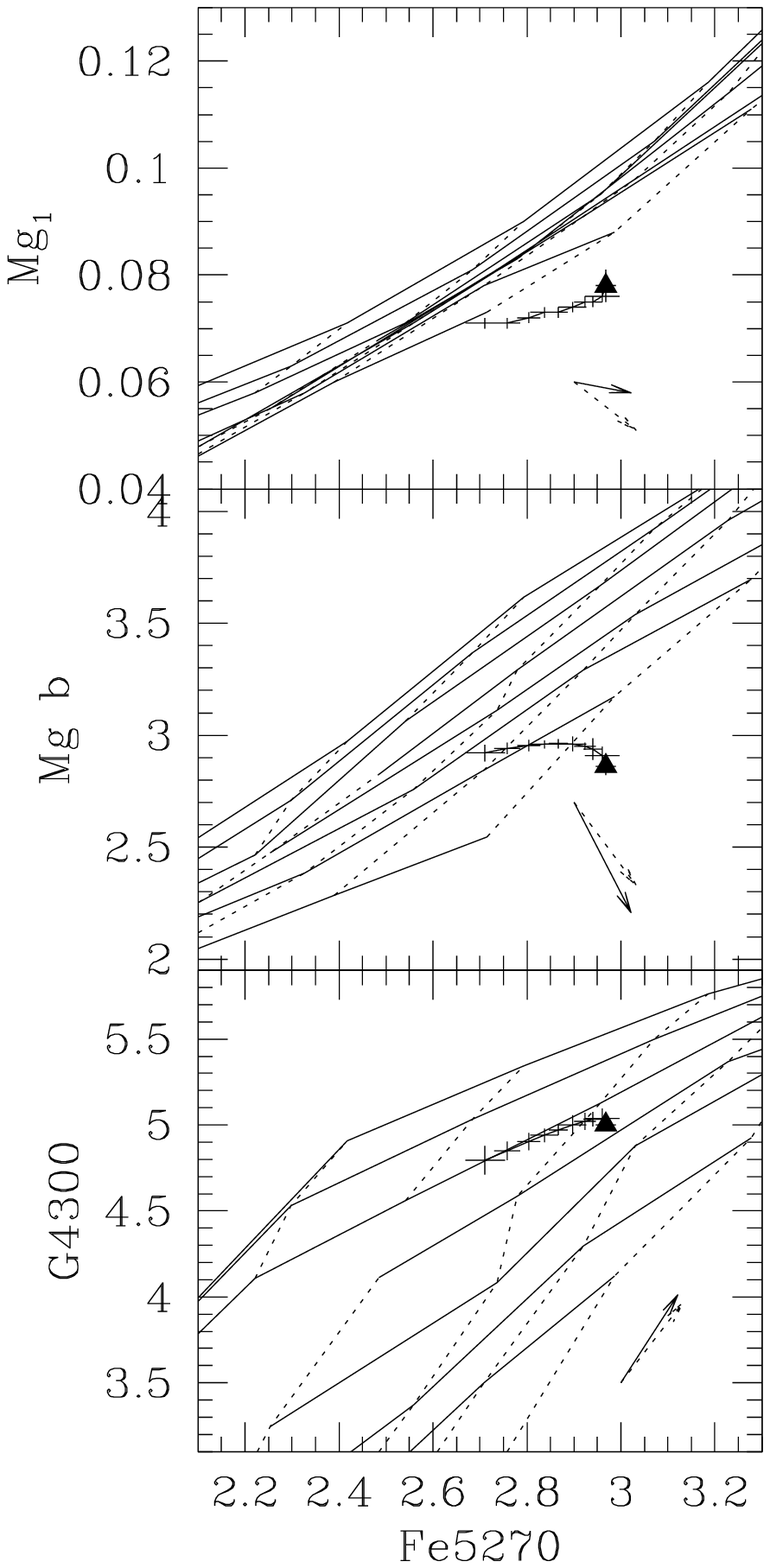}
\caption{ Index-index plots of the remaining Mg indices and
G4300. G4300 gives an age nearly identical to that of the Balmer
indices. None of the illustrated indices were used to iterate to the
final age plus abundance-pattern solution.
Symbols are as in Figure \ref{fig5}.
\label{fig9}}
\end{figure}

We illustrate the results by making a series of plots, shown in
Figures \ref{fig5}, \ref{fig6}, \ref{fig7}, \ref{fig8}, and
\ref{fig9}. The vectors drawn in the plots indicate the extent to
which the abundance ratio changes are needed to find a single mean age
for M32. The dotted vector is the displacement from the Table
\ref{tab2} nuclear observation and the adopted (age=4.75 Gyr,
[M/H]=0.02) model. The solid vector indicates the model displacement
if the abundance pattern of the previous paragraph is adopted. The
usually good agreement of the two vectors indicates that a single mean
age plus the adopted abundance pattern fits all indices to 0.43
$\sigma_{\rm IDS}$. This reduces the age scatter from the various
index-index diagrams from several gigayears to about $\pm$0.25 Gyr along an
(age, abundance) line segment from (5.0 Gyr, 0.00) to (4.5 Gyr,
0.05). This is an acceptable range, not a Gaussian $\sigma$.

Figure \ref{fig5} shows indices 
CN$_2$ (C- and N-sensitive), C$_2$4668 (C-sensitive), and Na D
(Na-sensitive, and also sensitive to interstellar absorption). 
CN$_2$ is high because of C, and N contributes negatively.
CN$_2$, as we have seen, has an interesting
near-nuclear dip so that the nucleus approaches the space between 3
and 5 Gyr marked by the models (landing in that spot would indicate
solar ratios). The inner $\sim10$ pc appear to have less N than the
bulk of the galaxy. Before we get too excited, a lot of the signal
goes away if the central pixel is omitted, and the central point has
atypically large errors. A prudent scientist should not generate
paragraphs of speculation based on 1 pixel. I therefore resist the
temptation to write of the possible influence of a central black hole
on N-rich ejecta of intermediate-mass AGB stars.

With the possible exception of nuclear N, the gradients follow the
overall age-metallicity trends sketched by the model grid. This is
also typical of Mg versus Fe [Figure \ref{fig6} and \citet{w98}]. This
simply means that the altered abundance ratios are approximately global
within the galaxy and more global than the age gradients and overall
abundance gradients.

Figure \ref{fig7} shows additional Fe-Fe plots.
Most of the
amplitude of the abundance vectors in these plots is caused by the
increase in C abundance. The Fe5335 prediction from \citet{houda}
models is quite different from that predicted by TB95. It is
of much interest to pursue the cause of this discrepancy, but at the
moment it is unknown and serves to indicate the extent to which it is
difficult to model Lick/IDS indices in synthetic spectra.
Figure \ref{fig8} shows indices affected by
elements N, C, and Na.
With Na D, there is an observational danger in that the Kitt
Peak sky contains Na D emission, so that Na D measurements in the
faint parts of the galaxy may suffer from sky-subtraction problems. I
mention this because, alone of the indices discussed, Na D fails to
span the full range of 0.25 dex from nucleus to 1 $R_e$.
Figure \ref{fig9} shows indices that were not fitted to the abundance
pattern.

% fig5 : x=fe5335, y=CN2,C4668,NaD
% fig6 : x=fe5335, y=Mg2,fe5709,Ca4227
% fig7 : x=fe5335, y=Fe4383,5270,5406
% fig8 : CN-4668,CN-NaD,4668-NaD
% fig9:  x=fe5270, y=Mg1, Mg b, G4300

\section{Summary}

Because of its high surface brightness, its proximity, the
availability of stellar photometry in the outer parts, and its
near-solar abundance ratios, M32 is an excellent laboratory for
testing stellar population theory and honing integrated light
models. In this paper I consider and discard the only long-slit
HST STIS spectra available because it proved impossible to clean
enough cosmic-ray hits from the images. Ground-based spectra from the MDM
Observatory, on the other hand, proved to be the best data set yet
assembled by a comfortable margin. In the MDM data, random
observational errors are much smaller than systematic errors.

The data reach slightly beyond 1 $R_e$ along the major axis of M32, an
angular distance of $44\farcs 4$, or a physical distance of 164
pc. Only the CN indices show much evidence of nonmonotonic
behavior by reaching a maximum about 4\arcsec\ from the nucleus and
decreasing both toward the nucleus and away from it. This may indicate
that the inner dozen pc of M32 are deficient in nitrogen, but
confirming evidence is needed to be sure about this. The only other
discontinuity present in M32, which has famously flat color gradients
except in the ultraviolet, is in the surface brightness profile at
2\arcsec\ radius. No echo of this profile break is seen in stellar
population indicators.

The mean age of the M32 nucleus, $\sim$4 Gyr, is consistent with other
studies. The mean age increases with radius to 8$-$10 Gyr at 1
$R_e$. Correction for abundance ratios ages the nucleus to 4.7
Gyr. The nuclear abundance is slightly supersolar but fades rapidly to
approximately [M/H]=$-$0.25 at 1 $R_e$. The differential radial trend
is more certain than the zero point.

Different from almost every other elliptical galaxy, M32 has [Mg/Fe] subsolar
in its dominant stellar population, about $-0.18$ dex. 
Other abundance results are [C/M]=+0.077, [N/M]=$-0.13$,
[Fe/M]$\approx$0.0, and [Na/M]=+0.12. Other elements considered by
TB95 do not show convincing evidence of enhancement: although the
fit slightly preferred [Cr/M]$=-0.15$ and
[Ti/M]$=-0.2$, these elements are consistent with [X/M]=0, along with
O, Ca, and Si, because the Lick indices are not very sensitive to
abundance changes in these elements.

Despite its astrophysical importance, oxygen cannot yet be
measured. Given the fact that all of the various age diagnostic
diagrams were splendidly self-consistent, oxygen is either near the
solar ratio or its abundance is not important for the structure of
metal-rich stars. If the former is true, then M32's abundance mixture is much
closer to the solar neighborhood than that of giant
elliptical galaxies. Nucleosynthetic theory has trouble predicting Na yields,
but large portions of the N and C abundance are thought to come from
mass loss in intermediate-mass stars. Using this as a handle, it seems
simple enough to invent schemes in which the timing of winds or the
presence of a black hole could alter the abundances of these common
elements. Future work should replicate the suspicious CN index drop
near the nucleus of M32. More flexible and internally consistent
stellar population models of the type proposed in \citet{w98} would
solidify the age and abundance ratio results presented here.

\acknowledgements 

The author gratefully acknowledges a suggestion from J. Kormendy to
correlate the light profile of M32 with stellar population indicators
and hopes he will forgive the negative result. J. J. Gonz\'alez kindly
provided his gradient data in electronic form. M. Houdashelt,
M. Briley, E. Baron, and A. Kunth each separately provided synthetic
spectra to confirm the TB95 overestimate of the effects of
carbon. This work was supported by grant HF-1066.01-94A awarded by the
Space Telescope Science Institute, which is operated by the
Association of Universities for Research in Astronomy, Inc., for NASA
under contract NAS5-26555; by Washington State University; and also by
the National Science Foundation under grant 0307487.

\end{document}